\documentclass[aps, amsmath, amssymb, superscriptaddress, nofootinbib, reprint,preprintnumbers]{revtex4-1}

\usepackage{graphicx}
\usepackage{epsfig}
\usepackage{epstopdf}
\usepackage{dcolumn}
\usepackage{bm}
\usepackage{subfigure}
\usepackage{xcolor} 
\usepackage{booktabs}
\usepackage{txfonts} 
\usepackage{lineno} 
\usepackage{bbold} 
\usepackage{amsmath}

\newcommand\SU{\mathrm{SU}}
\newcommand\U{\mathrm{U}}
\renewcommand\O{\mathrm{O}}
\newcommand\SB{\text{SB}}
\newcommand{\cL}{\ensuremath{\mathcal L} }
\newcommand{\cM}{\ensuremath{\mathcal M} }
\newcommand{\cO}{\ensuremath{\mathcal O} }
\newcommand{\psibar}{\ensuremath{\overline\psi} }
\newcommand{\vev}[1]{\ensuremath{\left\langle #1 \right\rangle} }
\newcommand{\EV}[2]{\ensuremath{\left\langle #1 \left| #2 \right| #1 \right\rangle} }

\newcommand{\Tr}{{\rm Tr}}

\begin{document}
\preprint{RBRC-1291}

\title{Linear Sigma EFT for Nearly Conformal Gauge Theories}

\author{T.~Appelquist}
\affiliation{Department of Physics, Sloane Laboratory, Yale University, New Haven, Connecticut 06520, USA}
\author{R.~C.~Brower}
\affiliation{Department of Physics and Center for Computational Science, Boston University, Boston, Massachusetts 02215, USA}
\author{G.~T.~Fleming}
\affiliation{Department of Physics, Sloane Laboratory, Yale University, New Haven, Connecticut 06520, USA}
\author{A.~Gasbarro}\email{andrew.gasbarro@yale.edu}
\affiliation{Department of Physics, Sloane Laboratory, Yale University, New Haven, Connecticut 06520, USA}
\affiliation{Physical and Life Sciences Division, Lawrence Livermore National Laboratory, Livermore, California 94550, USA}
\affiliation{Nuclear Science Division, Lawrence Berkeley National Laboratory, Berkeley, CA 94720, USA}
\affiliation{AEC Institute for Theoretical Physics, University of Bern, CH-3012 Bern, Switzerland}
\author{A.~Hasenfratz}
\affiliation{Department of Physics, University of Colorado, Boulder, Colorado 80309, USA}
\author{J.~Ingoldby}\email{james.ingoldby@yale.edu}
\affiliation{Department of Physics, Sloane Laboratory, Yale University, New Haven, Connecticut 06520, USA}
\author{J.~Kiskis}
\affiliation{Department of Physics, University of California, Davis, California 95616, USA}
\author{J.~C.~Osborn}
\affiliation{Computational Science Division, Argonne National Laboratory, Argonne, Illinois 60439, USA}
\author{C.~Rebbi}
\affiliation{Department of Physics and Center for Computational Science, Boston University, Boston, Massachusetts 02215, USA}
\author{E.~Rinaldi}
\affiliation{Nuclear Science Division, Lawrence Berkeley National Laboratory, Berkeley, CA 94720, USA}
\affiliation{RIKEN BNL Research Center, Brookhaven National Laboratory, Upton, New York 11973, USA}
\author{D.~Schaich}
\affiliation{AEC Institute for Theoretical Physics, University of Bern, CH-3012 Bern, Switzerland}
\author{P.~Vranas}
\affiliation{Physical and Life Sciences Division, Lawrence Livermore National Laboratory, Livermore, California 94550, USA}
\affiliation{Nuclear Science Division, Lawrence Berkeley National Laboratory, Berkeley, CA 94720, USA}
\author{E.~Weinberg}
\affiliation{Department of Physics and Center for Computational Science, Boston University, Boston, Massachusetts 02215, USA}
\affiliation{NVIDIA Corporation, Santa Clara, California 95050, USA}
\author{O.~Witzel}
\affiliation{Department of Physics, University of Colorado, Boulder, Colorado 80309, USA}
\collaboration{Lattice Strong Dynamics (LSD) Collaboration}


\begin{abstract}
We construct a generalized linear sigma model as an effective field theory (EFT) to describe nearly conformal gauge theories at low energies.
The work is motivated by recent lattice studies of gauge theories near the conformal window, which have shown that the lightest flavor-singlet scalar state in the spectrum ($\sigma$) can be much lighter than the vector state ($\rho$) and nearly degenerate with the PNGBs ($\pi$) over a large range of quark masses.
The EFT incorporates this feature. We highlight the crucial role played by the terms in the potential that explicitly break chiral symmetry.
The explicit breaking can be large enough so that a limited set of additional terms in the potential can no longer be neglected, with the EFT still weakly coupled in this new range.
The additional terms contribute importantly to the scalar and pion masses.
In particular, they relax the inequality $M_{\sigma}^2 \ge 3 M_{\pi}^2$, allowing for consistency with current lattice data.
\end{abstract}

\maketitle

\section{\label{sec:intro}Introduction}
In this paper we explore a linear sigma model as an effective field theory (EFT) description of gauge theories with approximate infrared conformal invariance.
Asymptotically free gauge theories exhibit conformal behavior in the IR when the number of fermions $N_f$ exceeds a critical value $N_f^c$.
When $N_f$ is taken just below $N_f^c$ the theory confines, but the low-energy physics below the confinement scale may be markedly different from QCD.
The EFT is motivated by recent work~\cite{Aoki:2014oha, Aoki:2016wnc, Appelquist:2016viq, Appelquist:2018yqe, Fodor:2014pqa, Fodor:2015vwa, Fodor:2016pls, Brower:2015owo, Athenodorou:2015fda, Appelquist:2014zsa} in which various nearly conformal gauge theories have been studied using lattice methods~\cite{DeGrand:2015zxa,Svetitsky:2017xqk}.
These theories, unlike QCD~\cite{Briceno:2016mjc, Liu:2016cba, Briceno:2017qmb, Guo:2018zss}, have been shown to possess a light flavor-singlet scalar state ($\sigma$) with mass similar to the pseudo-Nambu--Goldstone bosons (PNGBs or $\pi$), well separated from the vector meson ($\rho$) and other heavier resonances.\footnote{Throughout this work, we borrow the language of QCD to label hadron states.  For the lightest resonance in each channel, we denote the flavor-singlet scalar by $\sigma$, the flavor-singlet pseudoscalar by $\eta'$, the flavor-adjoint scalar by $a_0$, the flavor-adjoint pseudoscalar by $\pi$, the flavor-adjoint vector by $\rho$, and spin-1/2 baryon by $N$ or ``nucleon''.}

Light composite scalars have been reported in SU(3) gauge theory with eight flavors of fermions in the fundamental representation of the gauge group~\cite{Aoki:2014oha, Aoki:2016wnc, Appelquist:2016viq, Appelquist:2018yqe}, SU(3) gauge theory with two flavors in the symmetric (sextet) representation~\cite{Fodor:2014pqa, Fodor:2015vwa, Fodor:2016pls}, SU(3) gauge theory with four light and eight heavy fundamental flavors~\cite{Brower:2015owo}, and an SU(2) gauge theory with one adjoint flavor~\cite{Athenodorou:2015fda}.
This has motivated us to consider an EFT in which the flavor-singlet scalar is included as a dynamical degree of freedom.

Non-QCD-like confining gauge theories are interesting in their own right, and in addition these nearly conformal theories could be useful for constructing phenomenological models. In particular, the light $\sigma$ could be a viable candidate for a composite Higgs boson with dynamical electroweak symmetry breaking.
An important step towards developing such a model is to show that a confining gauge theory can generate a composite scalar state which is sufficiently light in the chiral limit. Currently, lattice calculations cannot be carried out close enough to the chiral limit to answer this question, but EFT techniques could shed light on this important issue. When the continuum EFT developed here is employed to fit lattice data, it must also be the case that any discretization effects are small.

Recent work indicates that chiral perturbation theory does not describe lattice data of SU(3) gauge theory with $N_f = 8$ fundamental flavors at currently accessible distances from the chiral limit~\cite{Appelquist:2014zsa, Gasbarro:2017fmi, Gasbarro:2017ccf}.
This is not surprising since the $\sigma$ is similar in mass to the pions in the quark mass regime studied, and so a perturbatively implemented EFT which omits the $\sigma$ resonance will not be an accurate description.
One can extend chiral perturbation theory to include the $\sigma$ by coupling a flavor-singlet scalar into the chiral Lagrangian in the most general way~\cite{Soto:2011ap, Hansen:2016fri}.
However, these models have a large number of low-energy constants and are difficult to constrain with limited lattice data.
There has been another effort to develop an EFT based on a hypothesis of spontaneously broken scale symmetry~\cite{Golterman:2016lsd, Golterman:2018mfm}, which has been shown to provide encouraging fits to the lattice data~\cite{Appelquist:2017wcg, Appelquist:2017vyy, Fodor:2017nlp}.

The linear sigma EFT considered here has attractive features in addition to accommodating a light flavor-singlet scalar. For example, lattice calculations of SU(3) gauge theory with eight fundamental flavors~\cite{Appelquist:2018yqe} indicate that the pion decay constant varies significantly with the bare quark mass. In the linear sigma model, where the scalar potential breaks chiral symmetry spontaneously, the pion decay constant has strong, tree-level dependence on the quark mass. Lattice calculations of the spectrum of SU(3) gauge theory with two sextet flavors~\cite{Fodor:2016pls} indicate that a multiplet of flavored scalar mesons ($a_0$) may become lighter than the vector mesons as the chiral limit is approached. Light flavored scalars are also predicted in Ref.~\cite{Kurachi:2006ej}. We include a multiplet of flavored scalars in the linear sigma EFT, but alternatively they can be removed from the spectrum by taking an appropriate limit.

In Section~\ref{sec:LSM} we introduce the linear sigma field, its transformation properties, and the leading order terms in the Lagrangian.
Since we are interested in applying the EFT to lattice computations necessarily carried out at nonzero quark mass, we include explicit chiral symmetry breaking terms in the Lagrangian.
In Section~\ref{sec:counting} we analyze the terms in the Lagrangian, ordering them depending on the size of the chiral symmetry breaking.
We develop an operator ordering rule to aid the analysis. We conclude that the chiral breaking can be large enough so that a limited set of terms in the potential becomes comparable to the one term that dominates in the small-quark-mass limit, with the EFT remaining weakly coupled in this new range.
The additional terms contribute importantly to the scalar and pion masses, relaxing  the inequality $M_{\sigma}^2 \ge 3 M_{\pi}^2$~\cite{KutiPresentation:2017}, allowing for consistency with current lattice data. We summarize our results in Section~\ref{sec:conc} and discuss open questions. In Appendix~\ref{appendixA}, we describe the limit that allows the flavored scalars to be removed from the spectrum, and in Appendix~\ref{appendixb} we discuss special considerations that apply to the case $N_f \le 4$.

\section{\label{sec:LSM}The Linear Sigma EFT}
An EFT is determined by the global symmetries of the system, a specification of the fields which transform according to some representation of the global symmetry group, and an ordering rule designating the relative importance of operators allowed by the symmetries.
For our case, the global symmetry of the EFT is determined by the underlying gauge theory, whose symmetry breaking pattern is $\SU_L(N_f) \times \SU_R(N_f) \times \U_V(1) \to \U_V(N_f)$ after the $\U_A(1)$ symmetry is broken explicitly at the quantum level.

We begin the construction of our EFT by taking the fields to transform in a linear multiplet of the flavor group. The linear sigma model for $N_f > 2$ was originally introduced by L\'evy~\cite{levy1967m} and further developed in much subsequent work in the context of three-flavor QCD, cf.~\cite{Schechter:1971qa, Schechter:1971tc}. Since the global symmetry group is a direct product group, the dynamical fields carry two indices, $M_a^{\overline b}$, where the unbarred subscript (barred superscript) transforms via linear action of a matrix in the fundamental (antifundamental) representation of $\SU_L(N_f)$ ($\SU_R(N_f)$).
\begin{equation}
  M_a^{\overline b} \to L_a^c M_c^{\overline d} \left(R^{\dag}\right)_{\overline d}^{\overline b}
\end{equation}
where $L, R \in \SU_{L, R}(N_f)$.
The field $M(x)$ transforms as a singlet under the $\U_V(1)$ symmetry, which we will disregard from here on.
Group indices will be suppressed in the remainder of the discussion.

When $N_f = 2$, the isometry $\SU_L(2) \times \SU_R(2) \sim \O(4)$ allows one to choose the linear multiplet to be real.
The four real degrees of freedom may be identified with three pseudoscalar pions and one scalar $\sigma$ which transform irreducibly in the adjoint and singlet representations of the unbroken $\SU_V(2)$, respectively. However, here we will work with a complex linear multiplet of scalars. When $N_f > 2$, the linear representation of $\SU_L(N_f)\times\SU_R(N_f)$ is necessarily complex. The 2$N_f^2$ real degrees of freedom may be identified with $N_f^2 - 1$ pseudoscalar pions and $N_f^2-1$ scalar $a_0$ states, each set transforming irreducibly in the adjoint representation of $\SU_V(N_f)$, as well as one pseudoscalar $\eta'$ and one scalar $\sigma$, each transforming as singlets under $\SU_V(N_f)$.

It is possible to express the complex matrix field $M(x)$ as a linear function of $2 N_f^2$ real component fields. However, we choose to use a nonlinear decomposition of $M(x)$ instead. This has advantages that we will make use of shortly.
\begin{equation}
  \label{eq:basis}
  M(x) = \exp\left[i\frac{\sqrt{N_f}}{F} \left( \frac{\eta'(x)}{\sqrt{N_f}}+ \pi^i(x) T^i\right) \right] \left(\frac{\sigma(x)}{\sqrt{N_f}} + a_0^i(x) T^i \right).
\end{equation}
The sum over the repeated adjoint indices is implied, and $T^i$ are the generators of $\SU(N_f)$ normalized such that $\Tr\left[T^i T^j\right] = \delta^{ij}$.
The mass scale $F$ is the vacuum expectation value (v.e.v.) of the $\sigma$ field.
Under parity, the matrix field transforms as $M(\vec{x},t) \to M^\dag(-\vec{x},t)$.

In the underlying gauge theory, the $\eta'$ degree of freedom is made heavy by mixing with topological fluctuations in the gluon field strength.
In Refs.~\cite{Meurice:2017zng, DeFloor:2018xrp}, the $\eta'$ degree of freedom is retained and a relationship between the $\eta'$ and $\sigma$ masses is derived.
Here, we manually remove the heavy $\eta'$ degree of freedom from the EFT by setting $\eta'(x) = 0$.
With $M(x)$ having been parametrized according to Eq.~(\ref{eq:basis}), the $\eta'$ degree of freedom is not mixed with the other field components under $\SU_L(N_f)\times \SU_R(N_f)$ chiral transformations.
So, when $\eta'(x)$ is set to zero, the fields in Eq.~(\ref{eq:basis}) still transform in a representation of the chiral symmetry (albeit a nonlinear one), but not of the $\U_A(1)$ symmetry.

Having removed the heavy $\eta'$, we could do the same with the $a_0$. There is some evidence that the $a_0$ is becoming lighter relative to the vector state as the fermion mass is reduced, so we will keep the $a_0$ in the EFT for our discussion. However, in current lattice data, the $a_0$ is still comparable in mass to the vector. It is possible to remove the $a_0$ from the EFT as explained in Appendix~\ref{appendixA}. We find that the tree-level masses of the other states and v.e.v. of the $\sigma$ field remain the same, even if the $a_0$ is taken out of the EFT.

In the leading, chirally symmetric part of the effective Lagrangian, we include all $\SU_{L}(N_f) \times \SU_{R}(N_f)$ invariant relevant and marginal operators.
For $N_f \le 4$, additional chirally invariant operators involving $\det M$ are marginal or relevant and also need to be included.
We consider this case separately in Appendix~\ref{appendixb}.
In what follows, we exhibit the terms appropriate for $N_f > 4$.
The effective Lagrangian is
\begin{align}
  \cL  = & \frac{1}{2} \Tr\left[\partial_\mu M \partial^\mu M^\dag\right] - V_0(M) - V_{\SB}(M), \label{eq:LLO}
\end{align}
where
\begin{align}
  V_0  = &\frac{-m_{\sigma}^2}{4} \Tr\left[M^\dag M\right] + \frac{m_{\sigma}^2 - m_a^2}{8 f^2} \Tr\left[M^\dag M\right]^2 \nonumber\\
   & + \frac{N_f m_a^2}{8 f^2} \Tr\left[M^\dag M M^\dag M\right]. \label{eq:v0}
\end{align}
$V_0$ is symmetric under the complete global symmetry group.
The potential $V_{\SB}$ represents the effects of the quark mass within the EFT, including the explicit breaking of $\SU_L(N_f) \times \SU_R(N_f)$ chiral symmetry.
We discuss the form of $V_{\SB}$ in Section~\ref{sec:counting}, keeping only the leading operators necessary to describe the mass spectrum of the EFT. For the analysis presented here, we work only to tree level.

We take $m_{\sigma}^2 > 0$, so that the theory exhibits spontaneous chiral symmetry breaking in the chiral limit.
We have parametrized the coefficients of the potential $V_0$ so that $m_\sigma$, $m_a$ and $f$ are the chiral-limit values of the mass of the $\sigma$ state, the mass of the $a_0$ states, and the v.e.v.\ of the field after spontaneous symmetry breaking.

The $M(x)$ field takes on a v.e.v.\ which is a global minimum of the potential.
We choose the v.e.v.\ to be oriented along the direction of the trace (the ``$\sigma$ direction'').
We denote quantities away from the chiral limit ($V_{\SB} \neq 0$) by capital letters: $M_{\sigma}$, $M_a$, $F$, and $M_\pi$ are the mass of the $\sigma$ state, the mass of the $a_0$ states, the v.e.v.\ of the field, and the mass of the $\pi$ states respectively.
$F$ is the same scale appearing in Eq.~(\ref{eq:basis}) in order to canonically normalize the pion kinetic term.
We expect that $f / F \ll 1$ for $F$ in the range of the current lattice data.

The minimum of the entire potential ($V = V_0+V_{\SB}$) is given by $\sigma = F$ and $a_0^i = \pi^i = 0$, with $F$ determined by the extremization condition $\delta V(M)/\delta M(x) = 0$, which reduces to
\begin{equation}
  \label{eq:minimum}
  \frac{F}{f}\left[\frac{F^2}{f^2} - 1\right] + \frac{2}{f m_\sigma^2}\left.\frac{\partial V_{\SB}}{\partial \sigma}\right|_{\sigma=F,\; \pi^i = a_0^i = 0} = 0.
\end{equation}
After reexpanding around this v.e.v., one arrives at the following expressions for the masses of the pions and scalars:
\begin{flalign}
  M_{\pi}^2    & = \left.\frac{\partial^2 V_{\SB}}{\partial \pi^{i\; 2}}\right|_{\sigma = F,\; \pi^i = a_0^i = 0}, \label{eq:LO1} \\
  M_{\sigma}^2 & = m_{\sigma}^2 \left( \frac{3}{2} \frac{F^2}{f^2} - \frac{1}{2}\right) + \left.\frac{\partial^2 V_{\SB}}{\partial \sigma^2}\right|_{\sigma = F,\; \pi^i = a_0^i = 0}, \label{eq:LO2} \\
  M_a^2        & = m_a^2 \frac{F^2}{f^2} + \frac{m_{\sigma}^2}{2} \left(\frac{F^2}{f^2} - 1\right) + \left.\frac{\partial^2 V_{\SB}}{\partial a_0^{i\; 2}}\right|_{\sigma = F,\; \pi^i = a_0^i = 0}. \label{eq:LO3}
\end{flalign}

We can derive the expression for the pion decay constant following the normalization conventions of Ref.~\cite{Pich:1995bw} by plugging the leading expression for the axial current, $A^{\mu\; i}(x) = (2 F/\sqrt{N_f}) \partial^\mu \pi^i(x)$, into the matrix element $\vev{0 \left| A^{\mu\; i}(0) \right| \pi^j(\vec{p})} = i \delta^{ij} \sqrt{2} F_\pi p^\mu$.
One finds
\begin{equation}
  F_\pi = \sqrt{\frac{2}{N_f}} F,
\end{equation}
where $F$ is the v.e.v.\ determined by the extremization condition Eq.~(\ref{eq:minimum}).
Here there is an important distinction between the linear sigma EFT and chiral perturbation theory.
In the latter, $F_{\pi}$ is a constant at tree level and depends on the explicit chiral breaking only at loop level.
In the linear sigma EFT, the pion decay constant depends on $V_{\SB}$ through Eq.~(\ref{eq:minimum}) at tree level.

\section{\label{sec:counting}Chiral Breaking}
The quark mass matrix is the source of explicit chiral symmetry breaking in the underlying gauge theory.
In the EFT, we take this breaking into account by introducing an auxiliary spurion field, $\chi(x)$, which transforms like $\chi(x) \to L \chi(x) R^\dag$ under a chiral rotation.
$V_{\SB}$ contains operators built out of $\chi(x)$ and $M(x)$ invariant under the chiral symmetry.
The symmetry is broken when the matrix field $\chi(x)$ is set to a constant value proportional to the quark mass matrix $\cM$,
\begin{equation}
  \label{eq:spurion}
  \chi(x) \to B \cM,
\end{equation}
where $B$ is a new low-energy constant with dimensions of mass.
The spurion then breaks the chiral symmetry in the EFT in the same way as the quark mass matrix in the underlying gauge theory.
We restrict to cases in which the quark masses are all degenerate, $\cM = m_q \mathbb{1}$.

The spurion construction catalogs the operators which may appear in $V_{\SB}$ on symmetry grounds, but we must determine the relative importance of these operators with respect to one another and with respect to operators that do not contain $\chi$.
With $V_{\SB} \neq 0$, the EFT is an expansion in $\partial/\Lambda$, $M(x)/\Lambda$, and  $\chi/\Lambda^2$, where $\Lambda$ is the EFT cutoff. With the EFT employed perturbatively, the cutoff can be taken no larger than the mass of the lightest excluded state. In the lattice data, for any value of $m_q$ in the current range, this is the vector state. We take $\Lambda$ to be of order this mass throughout the $m_q$ range.

Lattice data for nearly conformal theories~\cite{Aoki:2014oha,Aoki:2016wnc,Appelquist:2018yqe,Fodor:2016pls} indicates that\footnote{Throughout this work, we use the notation $A\sim B$ to denote $A=\cO(B)$.} $M_{\pi} \sim M_\sigma$ and $M_\sigma\sim F$ (the v.e.v.~of the field $\sigma(x)$). To ensure that the EFT reproduces the latter condition, we take the strength of the quartic potential to be $m_{\sigma}^2/f^2 \sim 1$, within the weak-coupling range. Finally, we  take the four momenta to be of order the particle masses, thus fixing $\partial/\Lambda \sim M_{\pi,\sigma}/\Lambda \sim F/\Lambda$. For the existing lattice data, the ratio $M_\sigma/\Lambda$ is not much smaller than $1/2$ but is tending to smaller values as $m_q$ decreases~\cite{Appelquist:2018yqe,Fodor:2016pls}. As for the $a_0$ mass $M_a$, we can set its value relative to $M_{\pi}$ and $M_{\sigma}$ by making an appropriate choice of $m_a^2$.

Next let us consider the order of magnitude of the chiral symmetry breaking expansion parameter. A measure of chiral symmetry breaking in the gauge theory is $m_q B_\pi$ where $B_\pi = \EV{0}{\psibar\psi}_{m_q = 0} / f_{\pi}^2$.
The v.e.v.\ is the chiral condensate for a single massless fermion flavor and $f_{\pi}$ denotes the chiral limit pion decay constant. In the chiral limit, we will have the GMOR relation~\cite{GellMann:1968rz} $(M_\pi^2 = 2 m_q B_{\pi})$ but away from the chiral limit, $2m_q B_\pi$ will not correspond directly to $M_{\pi}^2$, even at tree level.
We normalize the spurion so that $\chi = Bm_q\mathbb{1} = B_{\pi} m_q\mathbb{1}$ up to small corrections.

We compare the size of the chiral breaking effects to the ratio  $M_{\sigma} / \Lambda$, which controls the expansion in powers of fields and derivatives. In order to measure the size of $B m_q / \Lambda^2$ relative to $M_\sigma/\Lambda$, it is convenient to \emph{define} the quantity $\alpha$ by:
\begin{equation}
  \label{eq:oom}
  \frac{Bm_q}{\Lambda^2} = \left( \frac{M_{\sigma}}{\Lambda}\right)^\alpha.
\end{equation}
Each factor of $\chi/\Lambda^2$ in a Lagrangian operator therefore contributes a factor of $(M_\sigma/\Lambda)^{\alpha}$ to physical processes. We emphasize that $\alpha$ is not a free parameter, but rather is determined by $m_q$. As the chiral limit is approached, $\alpha$ becomes larger. In the limit $m_q \to 0$, Eq.~(\ref{eq:oom}) dictates that $\alpha \to \infty$.

We construct the EFT using the small quantities $\partial/\Lambda$, $M(x)/\Lambda$ and $\chi/\Lambda^2$. They have sizes
\begin{equation}
\frac{\partial}{\Lambda} \sim \frac{M(x)}{\Lambda} \sim \frac{M_\sigma}{\Lambda} \quad \text{and} \quad \frac{\chi}{\Lambda^2} = \left(\frac{M_\sigma}{\Lambda}\right)^\alpha\mathbb{1}\label{eq:oomparams}
\end{equation}
respectively.
To provide an estimate for the size of each operator coefficient, we take the Lagrangian to have the simple schematic form
\begin{equation}
  \label{eq:nda}
  \cL \supset \Lambda^4 \left(\frac{\partial}{\Lambda}\right)^{N_p}\left(\frac{M(x)}{\Lambda}\right)^{N_M} \left(\frac{\chi}{\Lambda^2}\right)^{N_{\chi}},
\end{equation}
such that the coefficient of an operator has order of magnitude $\Lambda^{4 - N_p - N_M - 2 N_{\chi}}$.
This form arises from dimensional analysis with the scale of each coefficient set by the mass ($\sim\Lambda$) of the lightest excluded state.
We note first that Eq.~(\ref{eq:nda}) sets the quartic couplings in $V_0$ to be $\cO(1)$.
This relatively weak value leads to $F \sim M_{\sigma}$ as seen in the lattice data.
Relative to the estimate of Eq.~(\ref{eq:nda}), however, the coefficient of the $\Tr\left[M^{\dag} M\right]$ operator in $V_0$ must be set to a smaller value (much less than $\Lambda^2$), a conventional tuning needed to produce the light scalar.

The quartic interaction as well as corrections to $V_0$ consisting of higher powers of $M(x)$ as dictated by Eq.~(\ref{eq:nda}) are relatively weak.
The same will be true of all the terms we employ in $V_{\SB}$.
Whether couplings with these sizes emerge from an underlying nearly conformal gauge theory with a relatively light scalar is an open question.
The answer will require further lattice study.
Here we assume that they do, at least for the operators that play a role here.

Using the order of magnitude estimate for the operator coefficients in Eq.~(\ref{eq:nda}) together with Eq.~(\ref{eq:oomparams}), one finds that each term in the Lagrangian has an order of magnitude size $M_\sigma^4 (M_\sigma / \Lambda)^{N_p + N_M + \alpha N_\chi - 4}$.
This motivates us to define a power counting dimension
\begin{equation}
  D = N_p + N_M + \alpha N_{\chi},
\end{equation}
where $\alpha$ is given by Eq.~(\ref{eq:oom}).
The leading-order Lagrangian is defined to include all terms with $D \le 4$.
In the chiral limit, taking $N_{\chi} = 0$, this corresponds to keeping only marginal and relevant operators. More generally, the terms to be included will depend on $\alpha$, that is, on the comparative size of chiral symmetry breaking. We will consider a relatively large amount of chiral symmetry breaking, corresponding to $\alpha$ as small as unity.

\subsection{The Breaking Potential}
\begin{table}[tbp]
  \centering
  \vspace{15 pt} 
  \renewcommand\arraystretch{1.6}  
  \addtolength{\tabcolsep}{3 pt}   
  \begin{tabular}{ |c|c| }
    \hline
    Symbol  & Operator                                                 \\
    \hline
    $\cO_1$ & $\Tr\left[\chi^\dag M + M^\dag \chi\right]$              \\
    $\cO_2$ & $\Tr\left[M^\dag M\right]\Tr\left[\chi^\dag M + M^\dag \chi\right]$          \\
    $\cO_3$ & $\Tr\left[(M^\dag M)(\chi^\dag M + M^\dag \chi)\right]$  \\
    $\cO_4$ & $\Tr\left[\chi^\dag M + M^\dag \chi\right]^2$                      \\
    $\cO_5$ & $\Tr\left[\chi^\dag \chi M^\dag M\right]$                          \\
    $\cO_6$ & $\Tr\left[\chi^\dag \chi\right]\Tr\left[M^\dag M\right]$                    \\
    $\cO_7$ & $\Tr\left[\chi^\dag M\chi^\dag M + M^\dag \chi M^\dag \chi\right]$ \\
    $\cO_8$ & $\Tr\left[\chi^\dag \chi\right]\Tr\left[\chi^\dag M + M^\dag \chi\right]$    \\
    $\cO_9$ & $\Tr\left[(\chi^\dag \chi)(\chi^\dag M + M^\dag \chi)\right]$      \\
    \hline
  \end{tabular}
  \caption{\label{table1}Operator content of the leading-order $(D\le 4)$ breaking potential when $\alpha = 1$, corresponding to a relatively large amount of chiral symmetry breaking. We show that this amount is required to fit currently available lattice data.}
\end{table}

For $\alpha$ as small as 1 and $N_f > 4$, the leading operators that enter the breaking potential are shown in Table~\ref{table1}.
The most general leading-order breaking potential may be parametrized as
\begin{equation}
  \label{eq:Vsb}
  V_{\SB} = -\sum_{i = 1}^9 \tilde c_i \cO_i(x).
\end{equation}
The first term (for which $D = 1 +\alpha$) takes the form $-\tilde c_1 B m_q \Tr\left[M + M^{\dag}\right]$. We set $\tilde{c}_1=f/\sqrt{N_f}$, ensuring that consistency with the GMOR relation near the chiral limit is maintained after having set $B=B_\pi$ up to small corrections. The appearance of $f$ in $\tilde{c}_1$ amounts to a tuning relative to the Eq.~(\ref{eq:nda}) estimate that this coefficient should be $\sim \Lambda$. This small value for $\tilde{c}_1$ also ensures that operators other than $\cO_1$ can play a significant role in $V_{\SB}$, even when $B m_q/\Lambda^2\ll1$. Since the coefficient of $\cO_1$ is symmetry protected, this value is technically natural. This is not true of the tuned coefficient of $\Tr\left[M^{\dag} M\right]$ described earlier. For all the other operators in Table~\ref{table1}, the principle of inclusion is that $D = 4$ when $\alpha = 1$. For each of these, we estimate the coefficients using Eq.~(\ref{eq:nda}).

We compute the leading-order expressions for the masses and the scalar v.e.v.\ (Eqs.~(\ref{eq:minimum}--\ref{eq:LO3})) for the general breaking potential, simplifying the expressions by absorbing factors of $N_f$ in the coefficients: $c_{2,9} = \sqrt{N_f}\tilde c_{2,9}$, $c_3 = \tilde c_3 / \sqrt{N_f}$, $c_{4,6} = N_f \tilde c_{4,6}$, $c_{5,7} = \tilde c_{5,7}$ and $c_8 = N_f^{3/2} \tilde c_8$.
We find
\begin{widetext}
  \begin{flalign}
    F^2 = & f^2 + \frac{2f^2}{m_{\sigma}^2} \left[ 2 B m_q \frac{f}{F} + 6 B m_q (c_2+c_3)F  + 2 B^2 m_q^2 (4 c_4 + c_5 + c_6 + 2 c_7) + 2 B^3 m_q^3 \frac{c_8+c_9}{F} \right], \label{eq:SB1} \\
    M_{\pi}^2 = & 2 B m_q \frac{f}{F} + 2 B m_q (c_2 + c_3) F + 8 B^2 m_q^2 (c_4 + c_7) + 2 B^3 m_q^3 \frac{c_8 + c_9}{F}, \label{eq:SB2} \\
    M_{\sigma}^2 = & m_{\sigma}^2 + 6 B m_q \frac{f}{F} + 6 B m_q (c_2 + c_3)F + 4 B^2 m_q^2 (4 c_4 + c_5 + c_6 + 2 c_7) + 6 B^3 m_q^3 \frac{c_8 + c_9}{F}, \label{eq:SB3} \\
    M_a^2 = & m_a^2 \frac{F^2}{f^2} + 4 B m_q \frac{f}{F} + 8 B m_q c_2 F + 2 B^2 m_q^2 (8c_4 + c_5 + c_6 + 2 c_7) + 4 B^3 m_q^3 \frac{c_8 + c_9}{F}. \label{eq:SB4}
  \end{flalign}
\end{widetext}
To ensure that $M_\sigma \sim F$ for all values of $Bm_q$, we set the strength of the quartic potential $m^2_\sigma/f^2\sim1$, a value within its weak-coupling range.

\subsection{\label{sec:regions}Large Quark-Mass Behavior}
Eqs.~(\ref{eq:SB2}, \ref{eq:SB3}) can be combined to express $M_{\sigma}^2$ in terms of $M_{\pi}^2$,
\begin{equation}
  \label{eq:msigrelation}
  3 M_\pi^2 - M_{\sigma}^2 + m_{\sigma}^2 = 4 B^2 m_q^2 (2 c_4 -c_5 -c_6 + 4 c_7),
\end{equation}
where $m_{\sigma}^2$ must be positive in a theory with underlying spontaneous symmetry breaking. For sufficiently small values of $B m_q$ (sufficiently large values of $\alpha$), the right hand side of Eq.~(\ref{eq:msigrelation}) will be highly suppressed and one finds the inequality $M_{\sigma}^2 \ge 3 M_{\pi}^2$, which is not respected by the lattice data.

This inequality is present quite generally for any $Bm_q$ small enough such that the operator $\cO_1$ dominates $V_{\SB}$. This can be seen from Eqs.~(\ref{eq:minimum}--\ref{eq:LO2}), the factor of $3$ in Eq.~(\ref{eq:LO2}) arising from the fact that the stabilizing potential is quartic. Although we have shown only that the inequality appears for $N_f > 4$, in Appendix~\ref{appendixb}, we show that it arises also for $N_f = 4$ and $N_f = 2$.

For large enough values of $B m_q$, however, this inequality is not in general present\footnote{The evasion of the inequality will depend on the signs as well as the order of magnitude of the coefficients $c_{4,5,6,7}$. These will be determined by details of the underlying gauge theory.}. To analyze the terms in Eqs.~(\ref{eq:SB1}--\ref{eq:SB3}) and determine the requisite size of $Bm_q$, we first use Eq.~(\ref{eq:nda}) to estimate the sizes of the operator coefficients:
\begin{align}
  c_{2,3} & \sim \Lambda^{-1} \nonumber\\
  c_{4,5,6,7} & \sim \Lambda^{-2}\\
  c_{8,9} & \sim \Lambda^{-3}.\nonumber
\end{align}
Using Eq.~(\ref{eq:oom}), the right hand side of Eq.~(\ref{eq:msigrelation}) can be estimated to be of order
\begin{equation}
  \label{eq:correction}
  4 B^2 m_q^2 (2 c_4 -c_5 -c_6 + 4 c_7) \sim M_{\sigma}^2 \left(\frac{M_\sigma}{\Lambda} \right)^{2\alpha - 2}.
\end{equation}
Thus if $\alpha$ can be taken as small as unity, that is if $Bm_q$ can be made as large as $M_{\sigma} \Lambda$, the unacceptable inequality cannot be established based only on order of magnitude estimates. We now consider more general features of the EFT for each of the qualitatively different regions of $Bm_q$.

\begin{figure}
  \centering
  \includegraphics[height=4.5cm]{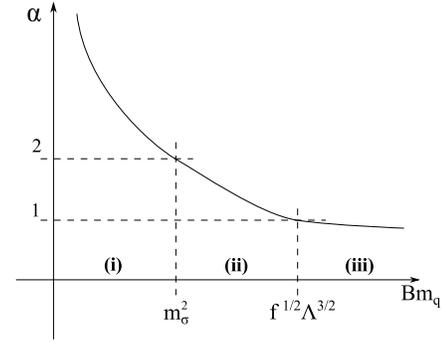}
  \caption{\label{Fig:cartoon}Cartoon showing the dependence of the quantity $\alpha$ (defined in Eq.~(\ref{eq:oom})) on the quark mass. Roman numerals refer to different quark mass regions described in the text below.}
\end{figure}

\vspace{8pt}
{\bf Region (i):} Near the chiral limit, where $Bm_q \lesssim m_{\sigma}^2 \sim f^2$, Eqs.~(\ref{eq:SB1}--\ref{eq:SB3}) lead to $M_{\pi}^2 \lesssim M_{\sigma}^2 \sim F^2 \sim f^2$, corresponding to $\alpha \gtrsim 2$ as shown in Fig.~\ref{Fig:cartoon}. The inequality $M_{\sigma}^2 \ge 3M_{\pi}^2$ is present.  Near the upper boundary of this region, the operator $\cO_1$ contributes at the same level as the terms in $V_0$, while the additional operators are suppressed.
\vspace{8pt}

{\bf Region (ii):} As $Bm_q$ is increased beyond $m_{\sigma}^2$, the quantities $M_{\pi}^2 $, $M_{\sigma}^2$ and $F^2$ begin to grow and the operator $\cO_1$ begins to dominate the $\sigma$ mass term in $V_0$. This is a transitional region. 
\vspace{8pt}

{\bf Region (iii):} Now suppose that $Bm_q$ becomes of order $f^{1/2}\Lambda^{3/2}$. This is achievable even if $\Lambda$, taken here to be of order the mass of the excluded vector state, increases moderately with $m_q$. We then have $M_{\pi}^2 \sim M_{\sigma}^2 \sim F^2 \sim f \Lambda$. This gives $Bm_q \sim M_{\sigma} \Lambda$ ($\alpha\approx1$), the requisite order of magnitude to avoid the inequality. At this level of chiral symmetry breaking, one can see that each Lagrangian operator in Table~\ref{table1} contributes at the same level ($\sim f^2 \Lambda^2$) as the dominant, quartic term in $V_0$. It can also be seen that symmetry-breaking operators with higher powers of $M$ or $\chi$, which have $D > 4$ when $\alpha = 1$, contribute at a lower level. 

The key condition $Bm_q \sim M_{\sigma} \Lambda$ maintains itself even as  $Bm_q$ is taken larger, into the range $f^{1/2}\Lambda^{3/2}< Bm_q \ll\Lambda^2$. As shown in Fig.~\ref{Fig:cartoon}, $\alpha$ stays close to 1 in this region. Here, Eqs.~(\ref{eq:SB1}--\ref{eq:SB3}) lead to $M_{\pi}^2 \sim M_{\sigma}^2 \sim F^2 \sim (Bm_q)^2 / \Lambda^2$, and each of the operators $\cO_{2-9}$ contributes to the Lagrangian at the same level $(Bm_q /\Lambda)^4$ as the quartic term in $V_0$. Operators with higher powers of $M$ or $\chi$ are suppressed to at least the level $(Bm_q)^{6}/\Lambda^8$.
\vspace{8pt}

To summarize, the key condition $Bm_q \sim M_{\sigma} \Lambda$ is met throughout the range $f^{1/2}\Lambda^{3/2} \lesssim Bm_q \ll \Lambda^2$, where the EFT remains within its range of validity. The picture here is analogous to the Banks-Zaks limit in the loop expansion of gauge theories. The leading $\cO_1$ term in $V_{\SB}$, being a relevant operator, has been assigned a relatively small coefficient. The other operators in Table~\ref{table1} can then contribute comparably, with additional operators being suppressed.

There is a price to pay for the relatively large amount of explicit chiral symmetry breaking invoked here.
The dependence of the EFT masses and the decay constant $F$ on the fermion mass is governed by Eqs.~(\ref{eq:SB1}--\ref{eq:SB4}).
When the $\cO_{2-9}$ terms become comparable to the leading terms, the form of this dependence becomes less evident.
Nevertheless, it should be possible to provide fits to the smooth, monotonic behavior of the lattice data~\cite{Aoki:2014oha, Aoki:2016wnc, Appelquist:2016viq, Appelquist:2018yqe, Fodor:2014pqa, Fodor:2015vwa, Fodor:2016pls}.
At tree level, the EFT coefficients entering these fits ($c_{2-9}$) depend on the scale setting scheme used to express the lattice data at each value of the quark mass. For a scheme in which $\Lambda$ varies with $m_q$, it will be necessary to model the quark mass dependence of $\Lambda$. There will also be quantum loop corrections to consider. While our EFT is relatively weakly coupled, with the strength of the quartic potential far smaller than $(4\pi)^2$, the size of these corrections will depend sensitively on $N_f$. Factors as large as $\cO(N_f^2)$ can enhance the loop factors of order $1 / (4\pi)^2$, lifting the effective loop expansion parameter closer to unity.

\section{\label{sec:conc}Conclusions}
We have constructed a generalized linear sigma model as an EFT for nearly conformal gauge theories with spontaneous chiral symmetry breaking.
Such an EFT may naturally accommodate several features indicated by recent lattice studies of these systems, including a light flavor-singlet scalar meson, significant dependence of the pion decay constant $F_{\pi}$ on the quark mass $m_q$~\cite{Aoki:2014oha, Aoki:2016wnc, Appelquist:2016viq, Appelquist:2018yqe, Fodor:2014pqa, Fodor:2015vwa, Fodor:2016pls, Brower:2015owo, Athenodorou:2015fda}, and the possibility of relatively light flavored scalars $a_0$~\cite{Fodor:2016pls,Kurachi:2006ej}.
It is also possible to remove the $a_0$ states from the spectrum by lifting their masses to infinity, as we explain in Appendix~\ref{appendixA}.

We investigated the linear sigma EFT by introducing a spurion field to represent the explicit breaking of chiral symmetry coming from the quark mass in the underlying gauge theory. This enabled the various chiral symmetry breaking operators to be enumerated and organized according to their order in an expansion in a chiral symmetry breaking parameter, proportional to the quark mass $m_q$.
To further facilitate the organization of operators, a measure $\alpha$ was employed, with values in the range $\alpha \gtrsim 2$ corresponding to the approach to the chiral limit. We presented the operator content of the leading-order chiral-breaking potential for $N_f>4$, summarized in Table~\ref{table1}. These operators become leading when the chiral breaking is larger, corresponding to the smaller value $\alpha \approx 1$.

We derived the tree level, leading-order expressions for the EFT quantities $F^2$, $M_{\pi}^2$, $M_{\sigma}^2$, and $M_a^2$, observing that the $\sigma$ mass and the $\pi$ mass are related by Eq.~(\ref{eq:msigrelation}). For small chiral breaking, the suppressed contributions of the operators $\cO_{4,5,6,7}$ imply that $M_{\sigma}^2 \ge 3M_{\pi}^2$, which is incompatible with the lattice results in Refs.~\cite{Aoki:2014oha, Aoki:2016wnc, Appelquist:2016viq, Appelquist:2018yqe, Fodor:2014pqa, Fodor:2015vwa, Fodor:2016pls, Brower:2015owo, Athenodorou:2015fda}. However, for sufficiently large chiral symmetry breaking, but still in a range where the EFT is under control (corresponding to $\alpha$ being close to unity), this inequality is relaxed leading us to conclude that the linear sigma EFT may indeed provide a viable description of these mass-deformed, nearly conformal gauge theories.

Looking to the future, it will be important to improve the lattice data for the $\SU(3)$ eight-flavor and other nearly conformal gauge theories, moving as close as possible to the chiral limit and minimizing lattice artifacts. One can then fit the data to Eqs.~(\ref{eq:SB1}--\ref{eq:SB4}), and estimate corrections arising at the quantum-loop level of the EFT. If favored by fits to lattice data, this EFT can be used to extrapolate the data to the chiral limit.

\section*{Acknowledgments}
We thank Maurizio Piai for helpful discussions. G.T.F.\ was supported by NSF grant {PHY-1417402}. A.H.\ and E.T.N.\ were supported by DOE grant {DE-SC0010005}; Brookhaven National Laboratory is supported by the DOE under contract {DE-SC0012704}.
R.C.B., C.R.\ and E.W.\ were supported by DOE grant {DE-SC0010025}.
In addition, R.C.B.\ and C.R.\ acknowledge the support of NSF grant {OCI-0749300}. A.G.\ acknowledges support under contract number {DE-SC0014664}. E.R.\ was supported by a RIKEN SPDR fellowship. P.V.\ acknowledges the support of the DOE under contract {DE-AC52-07NA27344} (LLNL). Argonne National Laboratory is supported by the DOE under contract {DE-AC02-06CH11357}.

\appendix
\section{\label{appendixA}Removing the Flavored Scalars}
The flavored scalar states $a_0$ do not affect the tree-level relations shown in Eqs.~(\ref{eq:SB1}--\ref{eq:SB3}), and these results continue to hold even if we remove the $a_0$ from the EFT by taking the limit $m_a \to \infty$.
In this limit, the fields are constrained to minimize the part of the potential proportional to $m_a$.
To derive this constraint, it is helpful to rewrite the potential shown in Eq.~(\ref{eq:v0}) as
\begin{align}
  V_0 = & \frac{m^2_a N_f}{8f^2}\Tr\left[\left(M^\dag M - \frac{1}{N_f}\Tr\left[M^\dag M\right] \mathbb{1}\right)^2\right] \nonumber\\
  & + \frac{m^2_\sigma}{8f^2}\left[\Tr\left[M^\dag M\right]-f^2\right]^2, \label{eq:v2}
\end{align}
where a constant has been added to $V_0$.
To minimize the first term in Eq.~(\ref{eq:v2}), the matrix that is squared and traced over should be set to zero, leading to a family of nonlinear constraints
\begin{equation}
  \label{eq:constraint}
  M^{\dag} M = \frac{1}{N_f} \Tr\left[M^{\dag} M\right] \mathbb{1}.
\end{equation}
Both sides of Eq.~(\ref{eq:constraint}) are hermitian, and taking the trace of each side does not lead to an independent constraint. Therefore, imposing Eq.~(\ref{eq:constraint}) as a constraint removes $N_f^2 -1$ real field degrees of freedom.

The constraint Eq.~(\ref{eq:constraint}) can be conveniently expressed in the field basis defined in Eq.~(\ref{eq:basis}),
\begin{equation}
  \frac{\sigma^2}{N_f}\mathbb{1} + 2\frac{\sigma}{\sqrt{N_f}} a_0^i T^i + a_0^i a_0^j T^i T^j = \frac{\sigma^2 + a_0^{i\; 2}}{N_f}\mathbb{1},
\end{equation}
independent of the $\pi$ and $\eta'$ fields.
The constraint is satisfied by setting $a_0^i = 0$.
The resulting Lagrangian with the constraint imposed is
\begin{equation}
  \label{eq:Lconstrained}
  \cL = \frac{\sigma^2}{2 N_f} \Tr\left[\partial_\mu \Sigma \partial^\mu \Sigma\right] + \frac{1}{2}\left( \partial_\mu\sigma\right)^2 + \frac{m_\sigma^2}{4} \sigma^2 - \frac{m_\sigma^2}{8 f^2} \sigma^4 - V_{\SB}|_{a_0^i = 0},
\end{equation}
where the pions are represented by the $\Sigma$ field satisfying the nonlinear constraint $\Sigma^{\dag} \Sigma = \mathbb{1}$.
Despite couplings in Eqs.~(\ref{eq:LLO}) and (\ref{eq:v0}) becoming large in the $m_a \to \infty$ limit, the resulting tree-level potential for the $\sigma$ Eq.~(\ref{eq:Lconstrained}) is weakly coupled as long as $m^2_\sigma/f^2$ is not too large.
The Lagrangian Eq.~(\ref{eq:Lconstrained}) is only leading order, and should be supplemented by higher-dimensional operators inherited from the linear sigma EFT.
In general, these operators will also be needed to cancel extra UV divergences that arise at loop level in the $m_a \to \infty$ limit.

\section{\label{appendixb}The Leading-Order Lagrangian for $N_f \le 4$}
In this appendix, we investigate whether for $N_f \leq 4$ the inequality $M_{\sigma}^2 \geq 3 M_{\pi}^2$ still arises for small $B m_q$, that is, whether it remains necessary to take $Bm_q$ large, into Region (iii) of Fig.~\ref{Fig:cartoon}. We show that this is the case for $N_f = 2$ and $4$. 

When $N_f \le 4$, operators that are invariant under $\SU_L(N_f)\times \SU_R(N_f)$ transformations involving $\det M$ become marginal or relevant and must be included in the leading-order Lagrangian. We shall consider only new operators that are invariant under the discrete parity symmetry $M(\vec{x},t) \to M^\dag(-\vec{x},t)$.

If the $\eta'$ state were included in the EFT, new determinant operators which break $\U_A(1)$ would provide it with mass. However in the following we manually remove the $\eta'$ from the EFT, as we did in Section~\ref{sec:LSM}. We first consider the simpler case of $N_f = 3$ or $4$, where only one new operator enters the chirally symmetric part of the potential $V_0(M)$. We then turn to the important case of $N_f = 2$.
\subsection{$N_f=3,\,4$}
In this case, $V_0(M)$ can be conveniently parametrized as
\begin{align}
  V_0(M) = & -\left[\frac{m^2_\sigma}{4}+\frac{N_f-4}{4N_f}\lambda f^2\right]\Tr\left[M^\dag M\right] \nonumber\\
           & + \left[\frac{m^2_\sigma-m_a^2}{8f^2}+\frac{\lambda}{8}\right]\Tr\left[M^\dag M\right]^2 \nonumber\\
           & + \left[\frac{m^2_a N_f}{8f^2}-\frac{\lambda}{4}\right]\Tr\left[M^\dag M M^\dag M\right] \nonumber\\
           & - \frac{\lambda f^4}{2N_f^2}\left[\frac{f}{\sqrt{N_f}}\right]^{-N_f} \left(\det M + \det M^\dag \right).
\end{align}
As before, we choose the parametrization such that the constants $f$, $m_\sigma$ and $m_a$ are the v.e.v.\ of the $\sigma$ field and the masses of the corresponding particles in the chiral limit.
This is the case for any value of the new dimensionless constant $\lambda$. With the new determinant operator included, Eqs.~(\ref{eq:minimum}--\ref{eq:LO3}) are now modified to
\begin{widetext}
\begin{align}
  0            = & \frac{F}{f}\left[\frac{F^2}{f^2} - 1\right]  + \frac{\lambda f^2}{N_f m_\sigma^2}\left[(N_f - 2)\frac{F^2}{f^2} - (N_f - 4)\frac{F}{f} - 2\left(\frac{F}{f}\right)^{N_f - 1}\right] + \frac{2}{f m_{\sigma}^2}\left.\frac{\partial V_{\SB}}{\partial \sigma}\right|_{\sigma = F,\; \pi^i = a_0^i = 0}, \label{eq:Fappend} \\
  M_{\pi}^2    = & \left.\frac{\partial^2 V_{\SB}}{\partial \pi^{i\; 2}}\right|_{\sigma = F,\; \pi^i = a_0^i = 0}, \label{eq:piappend}
  \end{align}
  \begin{align}
  M_{\sigma}^2 = & m_{\sigma}^2 \left[ \frac{3}{2} \frac{F^2}{f^2} - \frac{1}{2} \right] + \frac{\lambda f^2}{2 N_f}\left[3(N_f - 2)\frac{F^2}{f^2} - (N_f - 4)\frac{F}{f} - 2(N_f - 1) \left(\frac{F}{f}\right)^{N_f - 1}\right] + \left.\frac{\partial^2 V_{\SB}}{\partial \sigma^2}\right|_{\sigma = F,\; \pi^i = a_0^i = 0}, \label{eq:sigappend}\\
  M_a^2        = & m_a^2 \frac{F^2}{f^2} + \frac{m_{\sigma}^2}{2} \left(\frac{F^2}{f^2} - 1\right)  + \frac{\lambda f^2}{2 N_f}\left[(N_f - 6)\frac{F^2}{f^2} - (N_f - 4)\frac{F}{f} + 2\left(\frac{F}{f}\right)^{N_f - 1}\right] + \left.\frac{\partial^2 V_{\SB}}{\partial a_0^{i\; 2}}\right|_{\sigma = F,\; \pi^i = a_0^i = 0}.
\end{align}
\end{widetext}

We now test for the presence of the inequality when the chiral breaking is not large (below the threshold of Region (iii)), where $V_{\SB}$ is dominated by $\cO_1$. We find the mass relation
\begin{equation}
  \label{eq:massappend}
  M_{\sigma}^2 = m_{\sigma}^2 + 3 M^2_{\pi} + \lambda f^2 \frac{4 - N_f}{N_f} \left[\left(\frac{F}{f}\right)^{N_f - 2} - 1\right].
\end{equation}
For $N_f = 4$, the inequality is still present, meaning that $Bm_q$ must again be taken large, into Region (iii) to avoid the inequality. Intuitively, the new term in Eq.~(\ref{eq:massappend}) vanishes because the new determinant term in $V_0$ firstly preserves chiral symmetry (making no contribution to $M_{\pi}^2$) and secondly contributes to the $\sigma$ self interaction only a term of the form $\sigma^{4}$. 

For $N_f=3$, the inequality could be avoided even when only the operator $\cO_1$ is present, depending on the sign and size of $\lambda$. In this case, it would not be necessary to evade it by increasing $Bm_q$ and bringing the other operators of Table~\ref{table1} into the mix.

\subsection{$N_f=2$}
The $N_f = 2$ case is of particular relevance because the nearly conformal $\SU(3)$ gauge theory with $N_f=2$ sextet flavors has been studied on the lattice. Here, even more determinant operators are marginal and must be included in the leading-order Lagrangian. We first consider scalars transforming in the complex linear representation of $\SU_L(2)\times \SU_R(2)$. A full set of independent determinant operators to include in $V_0(M)$ is
\begin{gather*}
  \det M + \det M^{\dag}, \\
  \Tr\left[M^{\dag} M\right]\left[\det M + \det M^{\dag}\right].
\end{gather*}
The leading-order expressions for $M_\pi$, $M_\sigma$ and $F$ are unaffected by these operators and are given by Eqs.~(\ref{eq:minimum}--\ref{eq:LO2}). 
Therefore the inequality is also unaffected. This is because the determinant operators preserve chiral symmetry and contribute only quadratic and quartic $\sigma$ self-interactions. For scalars transforming in the real representation, there are no determinant operators that are independent. The inequality $M^2_\sigma \ge 3M^2_\pi$ continues to hold for quark masses in regions (i) and (ii) in this case too.

\bibliography{shortlinearsigma_APS}{}
\bibliographystyle{unsrt}
\end{document}